\documentclass[12pt]{article}

\usepackage{graphicx}
\usepackage{float}
\usepackage{cite}
\usepackage{amsfonts}
\usepackage{amssymb}
\usepackage{amsmath}
\usepackage{relsize}
\usepackage[compatibility=false]{caption}
\usepackage{subcaption}
\usepackage{xcolor}

\def\gtwid{\mathrel{\raise.3ex\hbox{$>$\kern-.75em\lower1ex\hbox{$\sim$}}}}
\def\ltwid{\mathrel{\raise.3ex\hbox{$<$\kern-.75em\lower1ex\hbox{$\sim$}}}}
\def\square{\kern1pt\vbox{\hrule height 1.2pt\hbox{\vrule width 1.2pt\hskip 3pt
   \vbox{\vskip 6pt}\hskip 3pt\vrule width 0.6pt}\hrule height 0.6pt}\kern1pt}

\begin{document}

\begin{titlepage}

\begin{flushright}
UFIFT-QG-26-04
\end{flushright}

\vskip 2.5cm

\begin{center}
{\bf Comments on Entire Functions of the Derivative Operator}
\end{center}

\vskip 1cm

\begin{center}
R. P. Woodard$^{1\dagger}$
\end{center}

\begin{center}
\it{$^{1}$ Department of Physics, University of Florida,\\
Gainesville, FL 32611, UNITED STATES}
\end{center}

\vspace{1cm}

\begin{center}
ABSTRACT
\end{center}
Many attempts to introduce fundamental nonlocality into quantum
(or classical) field theory are based on the assumption that 
exponentials of the d'Alembertian are positive-definite, so 
that these operators can be employed without engendering the 
Ostrogradskian instability associated with higher derivative
Lagrangians. {\bf This assumption is false.} Working in the 
simple context of a 1-dimensional, point particle $q(t)$, I 
demonstrate that the equation $\exp[T^2 \tfrac{d^2}{dt^2}] 
q(t) = 0$ has an infinite number of rapidly oscillating, 
exponentially rising and falling solutions. This infinite 
kernel is in one-to-one correspondence with the ability to 
specify ``initial value data'' {\it arbitrarily} over {\it 
any} finite interval $t_1 < t < t_2$. 

\begin{flushleft}
\end{flushleft}

\vspace{2cm}

\begin{flushleft}
$^{\dagger}$ e-mail: woodard@phys.ufl.edu
\end{flushleft}

\end{titlepage}

\section{Introduction}

One of Newton's greatest contributions was the injunction that
fundamental laws of physics should take the form of second order
differential equations. Whether or not one accepts this precept, 
the fact is that over three centuries of aggressive experimentation
and observation since the publication of {\it Principia} 
\cite{Newton:1687eqk} have failed to turn up even a single 
counterexample. For this to have been an accident stretches
credulity. Ostrogradsky's theorem  provides an explanation: 
theories whose Lagrangians contain nondegenerate higher time 
derivatives possess Hamiltonians which are linear in all but one 
of their canonical momenta \cite{Ostrogradsky:1850fid}. In an 
interacting, continuum field theory the vast entropy of phase 
space drives the decay rate to infinity \cite{Woodard:2015zca}. 
Because entire functions converge to their Taylor expansions,
the same conclusion applies to nonlocal models based on entire 
functions of the derivative operator \cite{Eliezer:1989cr}.

Those who accept that Newton got it right are frustrated by the
persistent belief in the viability of models with nondegenerate 
higher derivatives and analytic nonlocality. The impetus behind 
this seems to be the fact that general relativity is not 
perturbatively renormalizable \cite{tHooft:1974toh,Deser:1974cz,
Deser:1974cy,Deser:1974zzd,Deser:1974nb,Deser:1974xq,Goroff:1985sz,
Goroff:1985th,vandeVen:1991gw}. Stelle showed that including the
two quadratic curvature terms ($R^2$ and $C^2$) gives a 
perturbatively renormalizable theory \cite{Stelle:1976gc}. 
Stability poses no obstacle to adding the $R^2$ term because 
it induces dynamics in what was a negative energy, but completely
constrained, degree of freedom in general relativity 
\cite{Woodard:2006nt}. However, adding the $C^2$ term induces a 
massive, negative energy, spin two degree of freedom that makes 
the theory virulently unstable \cite{Woodard:2006nt}. If this 
instability were somehow not present, the problem of quantizing 
gravity would be solved \cite{Woodard:2009ns}. Hence the endless 
fascination with legitimizing this term.

Recent years have witnessed at least four distinct classes of efforts 
to evade the constraints imposed by Ostrogradsky's theorem:
\begin{enumerate}
\item{Skepticism about the reality of the classical instability, 
supported by numerical studies of discretized higher derivative 
models \cite{Deffayet:2025lnj,Deffayet:2023wdg};}
\item{Noncanonical quantizations of higher derivative models
\cite{Bender:2007wu,Bender:2008vh,Salles:2014rua,deOSalles:2018eon,
Donoghue:2019fcb,Donoghue:2019ecz,Mannheim:2020ryw,Donoghue:2021eto,
Holdom:2021hlo,Mannheim:2021oat,Liu:2022gun,Mannheim:2023mfp};}
\item{Insertion of nonlocal form factors in gravity 
\cite{Biswas:2013cha,Shapiro:2015uxa,Edholm:2016hbt,Buoninfante:2018xiw,
Koshelev:2024oms,Koshelev:2025pxg}; and}
\item{Inclusion of higher derivative and nonlocal interactions 
in studies of Asymptotic Safety \cite{Lauscher:2002sq,
Niedermaier:2006wt,Benedetti:2009rx,Donoghue:2019clr,Bonanno:2020bil}.}
\end{enumerate}
Regarding the first evasion, one should emphasize two points:
\begin{itemize}
\item{The virulence of the Ostrogradskian instability requires a 
model in which {\it continuum} negative and positive energy degrees 
of freedom {\it interact} \cite{Woodard:2006nt,Woodard:2015zca}. 
These are very safe assumptions for our physical universe, whose
dynamical variables are continuum fields which interact with 
gravity if nothing else. One should not be misled into doubting 
the Ostrogradskian instability by point particle models or models
for which interactions are absent.}
\item{Seeing the decay requires {\it generic} initial value data
\cite{Woodard:2006nt,Woodard:2015zca}. This is again a very safe 
assumption for the physical universe, each degree of freedom of 
which is compelled to undergo quantum fluctuations. The existence 
of long-lived classical solutions with the initial values of 
canonically conjugate pairs constrained more tightly than the 
uncertainty limit in no way invalidates the statement that the 
decay time is zero. The situation here is identical to that of 
the famous singularity theorems of general relativity 
\cite{Penrose:1964wq,Hawking:1965mf,Hawking:1970zqf}, which are 
perfectly consistent with solutions that do not form closed, 
trapped surfaces.}
\end{itemize}
Regarding the second evasion, the key point is that noncanonical 
quantizations forfeit the Correspondence Principle 
\cite{Mostafazadeh:2003tu,Mostafazadeh:2004mx,Mostafazadeh:2010yx}. 
That is, the classical limits of these models do not agree with 
the original theories. This might not pose any problem for a 
general system, which must be judged on the basis of its agreement 
with observed reality. However, it is fatal for gravity because 
everything we know about gravity at ordinary energies is consistent 
with the classical theory of a real-valued, generally coordinate 
invariant metric which obeys local field equations 
\cite{Will:2014kxa}. Noncanonical quantizations of higher 
derivative gravity cannot have these properties, so invoking these 
quantizations is akin to proclaiming the simple harmonic oscillator 
as ``quantum gravity''. This provides a wonderfully well-defined 
quantum theory, but it fails to predict the tides 
\cite{Woodard:2023tgb}.

The first two classes of evasions have been discussed above; this paper 
is devoted to a discussion of the 3rd and 4th classes. Researchers who
practice them typically concede the reality of the Ostrogradskian instability
for Lagrangians with finite numbers of nondegenerate higher time derivatives.
However, they hold out hope that form factors involving entire functions of
the derivative operator avoid the Ostrogradskian instability. Because entire 
functions converge to their Taylor series expansions, one would expect these 
models to contain an infinite number of higher derivative degrees of freedom, 
half of which are negative energy. Researchers who practice the 3rd and 4th
evasions deny this. For example, in the context of a point particle $q(t)$,
they assert that the most general solution of the equation,
\begin{equation}
\mathcal{O} \!\times\! [\ddot{q} + \omega^2 q] = 0 \qquad , \qquad 
\mathcal{O} \equiv \exp\Bigl[T^2 \tfrac{d^2}{dt^2}\Bigr] \; , \label{NLSHO}
\end{equation}
has only the two degrees of freedom of the simple harmonic oscillator,
\begin{equation}
q(t) = q_0 \cos(\omega t) + \tfrac{\dot{q}_0}{\omega} \sin(\omega t) \; .
\label{SHO}
\end{equation}
It is this belief I wish to address. In section 2 I will demonstrate that
equation (\ref{NLSHO}) actually possesses an infinite number of solutions,
in addition to (\ref{SHO}). The additional solutions have infinite 
frequency and are exponentially growing and falling, which cannot be 
dismissed out of hand because {\it it is the sort of problem predicted by
the Ostrogradskian instability.} In section 3 I show that this vast 
ensemble of additional solutions is associated with the ability to 
arbitrarily  prescribe $q(t)$ over any finite interval. My conclusions 
are given in section 4.

\section{The Kernel of $\mathcal{O}$}

Consider the equation,
\begin{equation}
\mathcal{O} q(t) = e^{-x^2} q(t) \; , \label{eigen}
\end{equation}
where $x$ is a real number. This is a linear differential equation with 
constant coefficients, so its general solution consists of a linear 
combinations of exponentials. Substituting $q(t) = e^{i\Omega t}$ in 
(\ref{eigen}) implies,
\begin{equation}
\exp\Bigl[-\Omega^2 T^2\Bigr] = e^{-x^2} \qquad \Longrightarrow \qquad
\Omega^2 = \tfrac1{T^2} \Bigl[x^2 + N \!\times\! 2\pi i \Bigr] \; , 
\label{Omsq} 
\end{equation}
where $N$ is an arbitrary integer. Solving for $\Omega$ gives,
\begin{equation}
\Omega = \tfrac{\pm 1}{\sqrt{2} \, T} \Biggl[ \sqrt{\sqrt{x^4 + 4 \pi^2 N^2}
+ x^2} + i \sqrt{\sqrt{x^4 + 4 \pi^2 N^2} - x^2} \Biggr] \; . 
\label{Omega}
\end{equation}

For any real $x$, we have found an infinite class of solutions, 
parameterized by $x$ and an arbitrary integer $N$, which oscillate 
with real frequency,
\begin{equation}
\vert{\rm Re}(\Omega)\vert = \tfrac{1}{\sqrt{2} \, T} 
\sqrt{\sqrt{x^4 + 4 \pi^2 N^2} + x^2} \; , \label{real}
\end{equation}
and rise or fall exponentially with rate,
\begin{equation}
\vert{\rm Im}(\Omega)\vert = \tfrac{1}{\sqrt{2} \, T} 
\sqrt{\sqrt{x^4 + 4 \pi^2 N^2} - x^2} \; . \label{imaginary}
\end{equation}
To recover the kernel of $\mathcal{O}$ we want to take $x$ to infinity.
It might be objected that this drives the oscillatory frequency 
(\ref{real}) to infinity as well and that these solutions are not
acceptable. This is a specious argument. The Ostrogradskian instability
manifests by imposing a bizarre time dependence on the dynamical 
variable and it is not valid to reject badly behaved functions and
then claim that there is no problem. One might equally well reject 
solutions of general relativity that develop closed, trapped surfaces
and then claim that the theory is free of singularities.

Another potential objection is that taking $x$ to infinity drives the 
exponential rate (\ref{imaginary}) to zero, so there are no longer an 
infinite number of solutions. However, one only gets zero rate if $N$ 
is fixed as $x$ goes to infinity. Because $N$ can be arbitrarily large,
it is possible to get near {\it any} rate $\frac{C}{\sqrt{2} \, T}$ by 
taking $N$ to obey,
\begin{equation}
N > \tfrac{1}{2\pi} \sqrt{2 C^2 x^2 + C^4} \; . \label{CN}
\end{equation}
The correct statement is therefore that even taking $x$ to infinity 
leaves an infinite number of solutions with distinct exponentially 
growing and falling rates (\ref{imaginary}).

Finally, it is worth noting that the factor of $e^{-x^2}$ in 
equation (\ref{eigen}) could easily be replaced with a general real 
number $\lambda$, which could even be negative,
\begin{equation}
\mathcal{O} q(t) = \lambda q(t) \; , \label{lambda}
\end{equation}
The previous derivation would go through with $x^2$ replaced by
$-\ln(\lambda)$, which would be complex for $\lambda < 0$. This 
observation has great significance because some proponents of 
nonlocal form factors concede that equation (\ref{lambda}) has an 
infinite collection of solutions for $\lambda \neq 0$, but try to
deny the existence of these solutions for $\lambda = 0$. That 
position becomes untenable in the face of solutions for either 
sign of $\lambda$. Solving (\ref{lambda}) for $\lambda < 0$ also
establishes that the operator $\mathcal{O}$ is not positive. Note
again that it is not valid to reject these solutions on the basis
that they lie outside of some nice set of functions, such as 
those which are square-integrable. The point of the Ostrogradskian
instability is that it engenders bad time dependence. One cannot
refuse to recognize the existence of bad time dependence and then 
proclaim that there is no instability.

\section{Arbitrary IVD for $\mathcal{O} q(t) = 0$ on $t_1 < t < t_2$}

The fact that there are an infinite number of linearly independent 
solutions to $\mathcal{O} q(t) = 0$ means that one can specify an 
arbitrary amount of initial value data (IVD). Naively, this means that the 
time dependence of $q(t)$ can be freely chosen over a finite interval. 
In this section I will prove that one can enforce $q(t) = f(t)$, for an 
arbitrary function $f(t)$, over any finite interval $t_1 < t < t_2$, by 
making a suitable choice for the behavior of $q(t)$ outside the interval.

To fix notation, let us suppose that $q(t) = 0$ for all $t < t_1$. The 
action of $\mathcal{O}$ on $q(t)$ can be represented as an integral 
transform,
\begin{equation}
\mathcal{O} q(t) = \tfrac1{\sqrt{4 \pi T^2}} \int_{t_1}^{\infty} \!\! 
dt' \exp\Bigl[-\tfrac{(t - t')^2}{4 T^2} \Bigr] \times q(t') \; . 
\label{transform}
\end{equation}
For any $t_1 < t < t_2$, we wish to show that $q(t')$ can be chosen for 
$t' > t_2$ so as to make,
\begin{equation}
\int_{t_1}^{t_2} \!\!\!\! dt' f(t') \exp\Bigl[-\tfrac{(t - t')^2}{4 T^2}
\Bigr] = -\int_{t_2}^{\infty} \!\!\!\! dt' q(t') \exp\Bigl[-\tfrac{(t - 
t')^2}{4 T^2}\Bigr] \; . \label{condition}
\end{equation}
Canceling a common factor and rearranging gives the simpler expression,
\begin{equation}
\mathcal{F}(t) \equiv \int_{t_1}^{t_2} \!\!\!\! dt' F(t') \!\times\! 
\exp\Bigl[\tfrac{t t'}{2 T^2}\Bigr] = -\int_{t_2}^{\infty} \!\!\!\! dt' 
Q(t') \!\times\! \exp\Bigl[\tfrac{t t'}{2 T^2}\Bigr] \; , 
\label{simpcon}
\end{equation}
where we define,
\begin{equation}
F(t) \equiv f(t) \!\times\! \exp\Bigl[-\tfrac{t^2}{4 T^2}\Bigr]
\qquad , \qquad Q(t) \equiv q(t) \!\times\! \exp\Bigl[-
\tfrac{t^2}{4 T^2}\Bigr] \; . \label{FQdef}
\end{equation}

That we can enforce (\ref{FQdef}) is almost obvious because $Q(t')$ 
can be freely chosen over an infinite interval to enforce conditions 
for $t$ in a finite interval. In fact, one can assume that $Q(t')$ 
vanishes after a finite interval of equal size. To be convinced, 
suppose we divide the interval $t_1 < t < t_2$ into $N$ segments of 
duration $\Delta t = \frac{(t_2 - t_1)}{N}$. Now define $t_n \equiv 
t_1 + (n - \frac12) \Delta t$, and consider the matrix equation,
\begin{equation}
\mathcal{F}(t_n) = -\sum_{m=1}^{N} \Delta t \, \exp\Bigl[
\tfrac{t_n t_m}{2 T^2}\Bigr] \!\times\! Q(t_m) \; . \label{matrix}
\end{equation}
Assuming the real, symmetric matrix $M_{n m} = \exp[
\frac{t_n t_m}{2 T^2}]$ has a nonzero determinant (which is easy to
verify for small $N$), the solution is straightforward,
\begin{equation}
Q(t_n) = -\sum_{m=1}^{N} M^{-1}_{n m} \!\times\! 
\tfrac{\mathcal{F}(t_m)}{\Delta t} \; . \label{solution}
\end{equation}
Even if the matrix $M_{nm}$ is not invertible, one can define a 
solvable problem by simply allowing $Q(t')$ to be nonzero slightly 
longer; there is an {\it infinite} range available. 

Two objections that might be raised are:
\begin{itemize}
\item{The equation $\mathcal{O} q(t) = 0$ is only being enforced
over the finite interval $t_1 < t < t_2$; and}
\item{The required future time dependence of $q(t')$ is bizarre, 
typically like $\exp[\frac{{t'}^2}{4 T^2}]$.}
\end{itemize}
The first objection can be overcome by using the same technique to
choose $q(t')$ even later to enforce $\mathcal{O} q(t)$ over the
first interval, and then repeating the process for each subsequent
interval. The second objection amounts to again denying the problem: 
it is {\it expected} for models with the Ostrogradskian instability 
to exhibit bizarre time dependence. One cannot forbid this and then
proclaim that there is no instability. 

\section{Conclusions}

In this short article I have considered the operator 
$\mathcal{O} \equiv \exp[T^2 \tfrac{d^2}{dt^2}]$, which is a point
particle analog of the form factors introduced in nonlocal models 
of gravity \cite{Biswas:2013cha,Shapiro:2015uxa,Edholm:2016hbt,
Buoninfante:2018xiw,Koshelev:2024oms,Koshelev:2025pxg}, and also 
the hoped-for fixed point of the Asymptotic Safety program 
\cite{Bonanno:2020bil,Knorr:2022dsx}. Researchers who pursue these 
programs assert that the solutions of the equation $\mathcal{O} 
[\ddot{q} + \omega^2 q] = 0$ are limited to those of the simple 
harmonic oscillator (\ref{SHO}). {\it This is false.} In section 2 
I demonstrated that there are an infinite number of additional 
solutions which oscillate with infinite frequency and grow or fall 
exponentially. In section 3 I demonstrate that these extra solutions 
are correlated with the ability to specify an infinite number of 
additional initial conditions, so that $q(t)$ can be chosen {\it 
arbitrarily} over {\it any} finite interval.

These are crushing problems, and they are precisely the sort of 
thing predicted by the Ostrogradskian instability. Inserting 
nonlocal form factors on propagators or vertices engenders this 
bizarre time dependence. It is not legitimate to dismiss the 
unwanted time dependence and then assert that there is no problem 
with nonlocal form factors. The correct conclusion is rather that 
the nonlocal models are unphysical. We all want a solution to the 
problem of quantum gravity, but failing to be clear-headed about
what does not work diverts attention from the search for truly 
viable solutions.

Two additional issues deserve comment. The first concerns the 
relation between the point particle model (\ref{NLSHO}) I have studied 
and the actual nonlocal quantum gravity models \cite{Biswas:2013cha,
Shapiro:2015uxa,Edholm:2016hbt,Buoninfante:2018xiw,Koshelev:2024oms,
Koshelev:2025pxg} proposed in the literature. Of course the dynamical 
variable of these models is the metric field, $g_{\mu\nu}(t,\vec{x})$, 
and their nonlocality extends to spatial derivatives as well as temporal 
ones. As an example, consider the model studied by Koshelev, Melichev 
and Rachwal \cite{Koshelev:2025pxg} in which the flat space perturbative 
field equation takes the form (cf. equations (A3-A7)),
\begin{equation}
m_{P}^2 \exp\Bigl[2 \omega(\tfrac{\square}{M_*^2})\Bigr] G^{\mu}_{~\nu} 
= \Bigl( {\rm Nonlinear\ Terms}\Bigr) \; , \label{KMReqn}
\end{equation}
where $G^{\mu}_{~\nu}$ is the Einstein tensor, $m_{P}$ is the Planck 
mass, $\square$ is the covariant d'Alembertian operator, $M_*$ is 
the energy scale of nonlocal effects, and $\omega(z)$ is an entire 
function. Because $\omega(\frac{\square}{M_*^2})$ is an entire function
of the derivative operator, it is asserted that the linearized
solutions (i.e., with the left hand side set to zero) are the same as 
those of general relativity. A moment's thought reveals that tensor
indices and spatial dependence of the dynamical variable change 
nothing about the central claim that exponentials of entire functions
of the derivative operator have zero kernel. One may as well examine
this claim in the context of a point particle as I have done.

The second issue concerns clarifying the respective function spaces 
typically assumed for nonlocal quantum gravity models and the larger 
function space which contains the extra solutions I have exhibited.
Explicit computations in these models are performed in Euclidean 
momentum space, which amounts to assuming that the temporal Fourier 
transform exists and can be analytically continued to Euclidean space.
There is no problem with this as long as one regards the introduction
of nonlocality as merely a tool for the ultraviolet regularization of
quantum field theory loop corrections and not as the basis for a new 
class of fundamental theories \cite{Evens:1990wf,Kleppe:1991rv}. 
Indeed, the technique has some promising features \cite{Kleppe:1990se}, 
and has even been pushed to 2-loop order \cite{Kleppe:1991ru}. 

The great question is whether or not the need for restrictions on the 
time dependence of dynamical variable renders a theory unphysical. 
There are different opinions about this. Some hold that the validity
of a physical assumption (such as the admissible space of solutions
of an equation) should be based on empirical evidence and mathematical
consistency. Others believe that one must be able to study a fundamental 
theory beyond perturbative loop corrections in Euclidean momentum space. 
For example, one must be able to study how nonlocal models affect the 
formation of singularities in gravity \cite{Biswas:2005qr}, and this 
requires abandoning arbitrary restrictions on the sort of time 
dependence which is permitted. As I have several times commented in this 
note, refusing to accept this would enable the absurd claim that general 
relativity is free of singularities because one can simply exclude 
singular solutions. One must instead permit the solutions of a truly 
fundamental theory to develop as they will, in the space of all functions.

\vskip .5cm

\centerline{\bf Acknowledgements}

I am grateful to A. Koshelev for conversations that motivated this
comment. This work was partially supported by NSF grant PHY-2207514 
and by the Institute for Fundamental Theory at the University of 
Florida.

\vskip .5cm

\centerline{\bf Note Added}

After the release of this work I was made aware of a paper by I. Shapiro
which also considers form factors involving exponentials of derivatives
\cite{Shapiro:2015uxa}.

\end{document}